\documentclass[conference]{IEEEtran}
\IEEEoverridecommandlockouts
\usepackage{cite}
\usepackage{amsmath,amssymb,amsfonts}
\usepackage{mathtools}
\usepackage{algorithm, algorithmic}
\usepackage{graphicx}
\usepackage{textcomp}
\usepackage[numbers,sort&compress]{natbib}
\usepackage{caption}
\usepackage{balance}
\usepackage{cleveref}
\usepackage{color}
\usepackage{url}
\usepackage{float}
\usepackage{enumitem}
\usepackage{svg}
\usepackage{textcomp}
\usepackage{xcolor}
\usepackage{subcaption}
\usepackage{tabularx}
\usepackage{xurl, balance}
\usepackage{tabularray}

\def\BibTeX{{\rm B\kern-.05em{\sc i\kern-.025em b}\kern-.08em
    T\kern-.1667em\lower.7ex\hbox{E}\kern-.125emX}}
\begin{document}

\title{Benchmarking Toxic Molecule Classification using Graph Neural Networks and Few Shot Learning.\\}

\author{
\IEEEauthorblockN{Bhavya Mehta\IEEEauthorrefmark{1}\textsuperscript{\textsection},
Kush Kothari\IEEEauthorrefmark{2}\textsuperscript{\textsection}, Reshmika Nambiar\IEEEauthorrefmark{3}\textsuperscript{\textsection} and
Seema Shrawne\IEEEauthorrefmark{5}}
\IEEEauthorblockA{Department of Computer Engineering and Information Technology,
Veermata Jijabai Technology Institute\\
Mumbai, India.\\
Email: \IEEEauthorrefmark{1}bdmehta\_b19@ce.vjti.ac.in,
\IEEEauthorrefmark{2}kmkothari\_b19@ce.vjti.ac.in,
\IEEEauthorrefmark{3}rsnambiar\_b19@ce.vjti.ac.in,
\IEEEauthorrefmark{5}scshrawne@ce.vjti.ac.in}}

\maketitle
\begingroup\renewcommand\thefootnote{\textsection}
\footnotetext{Equal contribution}
\endgroup

\begin{abstract}
Traditional methods like Graph Convolutional Networks (GCNs) face challenges with limited data and class imbalance, leading to suboptimal performance in graph classification tasks during toxicity prediction of molecules as a whole. To address these issues, we harness the power of Graph Isomorphic Networks, Multi Headed Attention and Free Large-scale Adversarial Augmentation separately on Graphs for precisely capturing the structural data of molecules and their toxicological properties. Additionally, we incorporate Few-Shot Learning to improve the model's generalization with limited annotated samples. Extensive experiments on a diverse toxicology dataset demonstrate that our method achieves an impressive state-of-art AUC-ROC value of 0.816, surpassing the baseline GCN model by 11.4\%. This highlights the significance of our proposed methodology and Few-Shot Learning in advancing Toxic Molecular Classification, with the potential to enhance drug discovery and environmental risk assessment processes.
\end{abstract}

\begin{IEEEkeywords}
Graph Neural Networks, Graph Isomorphic Network, Multi Headed Attention, Graph Data Augmentation, Few Shot Learning, Toxicity Prediction.
\end{IEEEkeywords}

\section{Introduction}\label{sec1}
Toxicological assessment of molecular compounds plays a pivotal role in drug discovery, environmental risk assessment, and chemical safety evaluation. Accurate prediction of a molecule's toxicity is crucial in ensuring the development of safe and effective drugs while minimizing potential harm to both human health and the environment.

 Traditional methods of toxic molecule detection [\cite{tradmethod1},\cite{tradmethod2}] possess some inherent limitations. This is because conducting experiments to synthesize a compound and then analyzing its toxicity is time-consuming and often very expensive. It consumes a lot of resources and is not feasible for large-scale testing of molecules.

A number of approaches based on machine learning have also recently been proposed. The methods described above use several molecular characteristics, such as their physical and chemical properties, to predict their toxicity. However, a present problem in the field is lack of sufficient labelled data, due to the difficulties faced in synthesizing and testing new molecules, as explained above. Moreover, often these machine learning techniques only look at certain numerical properties of the molecules and fail to take into consideration the structural aspects of the molecule.

In recent literature, a lot of research is being done in representing molecules as graphs and processing them through Graph Neural Networks (GNNs). While this method does not tackle the low-data scenario we often face in toxicity prediction, newer methods have integrated few-shot learning into GNNs, like the Adaptive Step Model-Agnostic Meta-Learner (AS-MAML). We believe that this intersection of graph-embedding algorithms and few-shot learning is key to creating effective models for molecular toxicity prediction.

The research problem addressed in this paper is to investigate and propose enhancements to the GNN-specific few-shot learning technique in order to achieve favorable results in the toxicity prediction task on the Tox21 data set under the few shot learning scenario.

\section{Background}\label{sec2}
Before delving into the specifics of the architecture, it is essential to provide some background information that will be helpful for better understanding.

\subsection{Few Shot Learning (FSL) }
For quite a few years now, FSL \cite{fsl2} has been an important topic being researched in the field of machine learning. As suggested by \citeauthor{fsl1} in \cite{fsl1}, it is the ability of an algorithm to generalize well from limited data points with supervised information available for every class. 


To achieve this, we employ Model-Agnostic Meta-Learning (MAML) given by \citeauthor{maml} which aims to find a good initialization for the model parameters $\theta$, for rapid adaptation to novel classes with only a few labeled examples. This is done by optimizing the model's performance on a set of meta-training experiments, where each task simulates a few-shot learning scenario.

\subsection{Adaptive Step Model Agnostic Meta Learning (AS-MAML) }
Introduced by \citeauthor{asmaml} in \cite{asmaml}, it is a meta-learning technique that builds upon the MAML\cite{maml} algorithm by introducing an Adaptation Controller that employs reinforcement learning techniques to determine the optimal step size and when to stop the adaptation process. A StopController model, incorporating LSTM \cite{lstm} layers and a sigmoid function, estimates the probability of stopping the adaptation process based on the training loss and embedding quality. This addresses the challenge of finding the optimal learning rate and step size in MAML-based meta-learning approaches.





\subsection{Graph Convolution Network (GCN) }
GCNs were introduced as a way to extend convolutional neural networks (CNNs) to handle irregular and non-Euclidean data, such as molecular structures, recommendation systems etc where traditional grid-like data structures like images or sequences do not apply. \citeauthor{gcn} in \cite{gcn} mentions that the key challenge in processing graph data is that the number of nodes and their connectivity can vary widely from one graph to another. GCNs address this challenge by learning to exploit the local neighborhood information of each node in the graph to make predictions. The core idea behind GCNs is to perform node feature aggregation through a series of graph convolutions, enabling nodes to gather information from their neighbors and incorporate it into their own representations. Mathematically, the message passing operation in a GCN can be represented as

\begin{equation}\label{eq:example}
   F^{(i+1)} = \sigma \left( \hat{D}^{-\frac{1}{2}} \hat{A} \hat{D}^{-\frac{1}{2}} F^{(i)} W^{(i)} \right)
\end{equation}

Here, \\
- \( F^{(i+1)} \): Feature matrix of all nodes at layer \( i+1 \).\\
- \( \sigma \): Activation function (Non -linear)\\
- \( \hat{A} \): The graph adjacency matrix with added self-loops.\\
- \( \hat{D} \): Diagonally arranged node degrees.\\
- \( H^{(i)} \): The feature matrix of all nodes at layer \( i \).\\
- \( W^{(i)} \): Layer \( i \)'s learnable weight matrix.\\

The GCN is capable of capturing increasingly complicated linkages and higher-order dependencies inside the graph by stacking numerous graph convolutional layers. The final node representations can be used for plenty of subsequent tasks, such as node classification, link prediction, and graph classification, subsequent to the application of numerous layers.

\subsection{Graph Isomorphic Network (GIN) }
Graph Isomorphic Networks by \citeauthor{gin} in \cite{gin} are a class of deep learning models designed for graph classification tasks. Unlike traditional GCNs, GINs do not rely on graph structure during message passing, making them more flexible and suitable for various graph types. The core idea behind GINs is to employ an aggregation function that is permutation-invariant to the node ordering, ensuring that the model produces the same output regardless of how the nodes are arranged. This property allows GINs to capture the global graph information effectively and provide more robust representations for graph classification tasks. Mathematically, the GIN update rule for a node \(v\) at layer \(l\) can be defined as: 

\begin{equation}\label{eq:example}
f_v^{(i+1)} = \text{MLP}^{(i)} \left( \sum_{u \in \mathcal{N}(v)} f_u^{(i)} + (1+\epsilon^{(i)}) \cdot f_v^{(i)}  \right)
\end{equation}

where \(f_v^{(i)}\) is the feature representation of node \(v\) at layer \(i\), \(\mathcal{N}(v)\) represents the neighbors of node \(v\), \(\text{MLP}^{(i)}\) is a Multi-Layer Perceptron applied element-wise, and \(\epsilon^{(i)}\) is a learnable parameter that allows GINs to learn and adapt the importance of self-information during aggregation.

Overall, Graph Isomorphic Networks have proven to be effective and efficient for graph classification tasks, offering a powerful approach to learn expressive and permutation-invariant graph representations that can handle various graph structures with high accuracy.

\subsection{Free Large-scale Adversarial Augmentation on Graphs (FLAG) }
It is a technique for enhancing graph data to imprive GNNs' performance. FLAG by \citeauthor{flag} in \cite{flag} suggests augmenting node properties rather than modifying graph topological structures, which is where the majority of existing graph regularizers concentrate their efforts. It improves generalization to out-of-distribution samples by iteratively enhancing node characteristics with gradient-based adversarial perturbations during training. This makes the model invariant to tiny fluctuations in input data. This all-purpose method called FLAG is effective for tasks involving graph classification, link prediction, and node classification. Adversarial data points are created and then inserted into the training data as part of the adversarial training process. The objective of this min-max optimization problem is to minimize the objective function while keeping the perturbation within a predetermined bound.

\section{Related Works}\label{sec3}

Some of the earliest works in toxicity prediction include DeepTox by \citeauthor{deeptox}, who used chemical properties of these compounds fed into a Deep Neural Network to predict their toxicity. By using this method and ample of labelled data \citeauthor{deeptox}, manages to achieve an 0.92 AUC value. \citeauthor{allsmilesvae} introduced All SMILES VAE \cite{allsmilesvae}, a generative model which uses variational autoencoders (VAEs) for generating SMILES strings using stacked RNNs. The model surpassed state-of-the-art methods and achieved an ROC-AUC score of 0.871 on the dataset. Censnet by \cite{censnet} learns node and edge features through the use of novel propagation rules while switching the roles of nodes and edges. The method attains about 0.79 AUC score at most on the Tox21 dataset when tested under various splitting scenarios. \citeauthor{unimol} proposed Uni-Mol \cite{unimol}, a framework that incorporates the pretraining of transformers in order to use 3D information. It was evaluated on Tox21 as a downstream task and outperformed several methods, achieving an ROC-AUC score of 0.796. 

Graph Multiset Transformer (GMT) \cite{gmp} adopts a novel pooling method wherein multi-head attention is used for learning node interaction based on task relevance. An AUC score of about 0.773 was obtained on Tox21. Meta-MGNN by \citeauthor{metamgnn} employs meta-learning to learn molecular representations under few-shot settings. It uses pretrained GNNs and leverages additional tasks to be optimised. When tested on Tox21, an AUC score of 0.769 was obtained under the one-shot setting and about 0.78 under the five-shot setting, outperforming several baseline models. However very few works have obtained significant results in few shot domain with graphs.\citeauthor{semisupgnn} in \cite{semisupgnn} achieves an average ROC-AUC score of 0.757 employing the Mean Teacher Semi-Supervised ML Algorithm, which is a 6\% increase over GCN models trained using supervised and conventional ML techniques. However for low data scenarios, very few works have been able to get significant results.

\section{Tox21 Dataset}\label{sec4}
Tox21 is a dataset containing measurements of toxicity of 12 thousand molecules against 12 target proteins. It aims to help analyse the performance of models in predicting the biochemical activity of compounds using their chemical structure. We use the AhR sub-dataset from Tox21 that focuses on chemicals' interactions with this Aryl hydrocarbon Receptor, a ligand-activated transcription factor that is essential for the toxic response to toxins and medications. The dataset is open source and can be downloaded from Tox21 AHR\footnote{\url{http://bioinf.jku.at/research/DeepTox/tox21.html}}.

Each chemical compound in this collection is represented as a graph, with atoms serving as nodes and chemical bonds between atoms serving as edges. Molecules' structural information is preserved in the graph representation, making it ideal for GNN-based approaches that can efficiently handle graph-structured data. By learning from the graph structure and associated node features, GNNs can discern complex relationships and identify key structural characteristics associated with toxic and non-toxic compounds.



\section{Baseline Model}
\begin{figure}[t]
    \centering
    \includegraphics[width=3in]{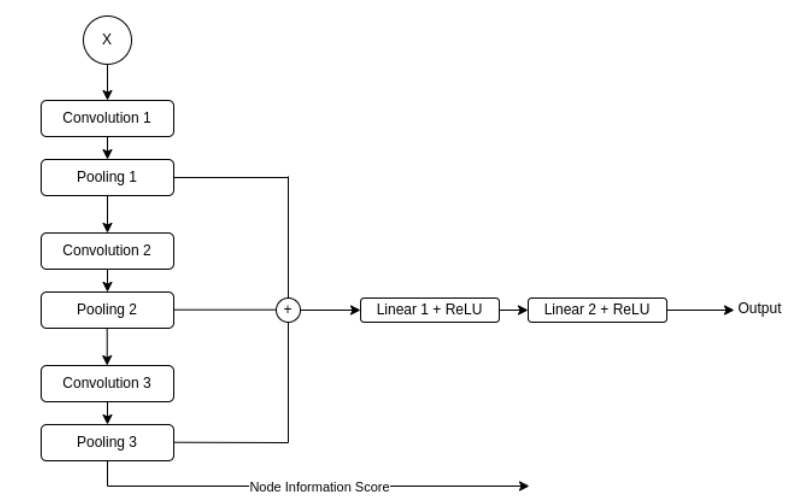}
    \caption{The image illustrates the baseline sub-architecture.The resulting output vector is subsequently fed into a binary sigmoid classifier. The obtained Node Information Score is utilized by the FSL Reinforcement Learning Agent to optimize gradients and weights, thereby achieving a faster convergence rate.}
    \label{fig:baseline}
\end{figure} 
The initial configuration we are evaluating serves as the baseline, which is the standard GCN + AS-MAML model utilizing the few-shot learning setup detailed earlier. While we remain consistent with the framework described in the paper, there is one notable difference: we do not employ distinct classes for training and validation. This configuration consists of three successive layers: a GCN convolution layer, followed by a TopK Pooling Layer \cite{topk}, each with a hidden layer dimension of 128. The Baseline Architecture is shown in Fig.\ref{fig:baseline}, has a validation accuracy of 65.02\% and an AUC-ROC value of 0.732 on Tox21 AhR data.

\section{Proposed Architectures}\label{sec5}
In this research paper, we introduce and empirically evaluate three distinct architectural frameworks, each of which outperforms the baseline model in terms of achieved results. These three novel architectures systematically introduce variations across distinct components of the baseline model's structure the body, the input and the output, consequently enhancing the model's capacity to capture intricate patterns and further enriching its learning capabilities.

\subsection{Augmenting Input data using FLAG}
\subsubsection{Architecture}
\begin{figure}[t]
    \centering
    \includegraphics[width=3in]{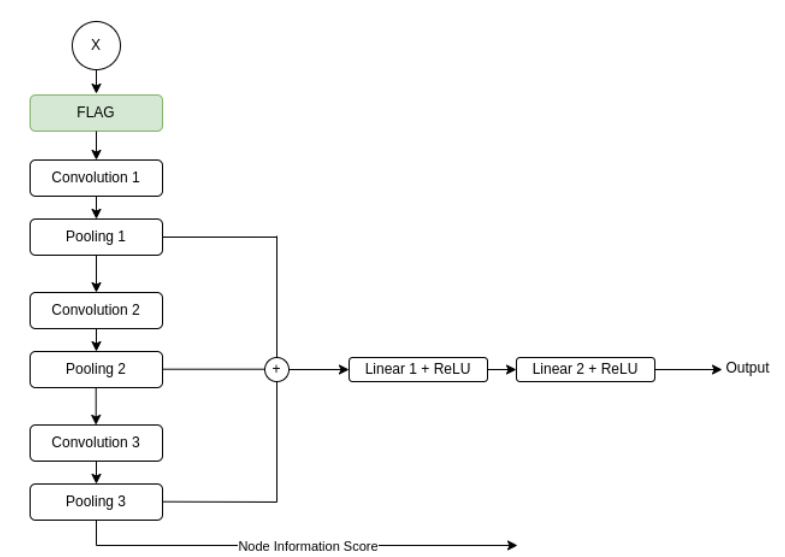}
    \caption{Proposed sub-architecture of FLAG+GCN based classification model.}
    \label{fig:FLAG_arch}
\end{figure} 

The first suggested setting adds a preprocessing step of FLAG in order to augment the data being fed into the model as shown in Fig.\ref{fig:FLAG_arch}. This adds perturbations to node features and provides greater variations in novel tasks available for few-shot learning. The aim of FLAG is to generate additional realistic graph instances that maintain the underlying distribution of the original data, effectively expanding the dataset and boosting model generalization.

\subsubsection{Experimental Results}
\begin{figure}[t]
    \centering
    \includegraphics[width=3in]{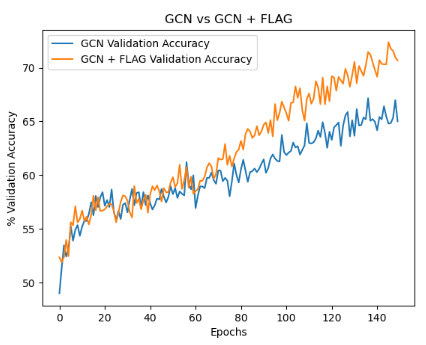}
    \caption{A comparision of validation accuracy for GCN vs GCN+Flag method.}
    \label{fig:FLAG_acc}
\end{figure} 
\begin{figure}[t]
    \centering
    \includegraphics[width=3in]{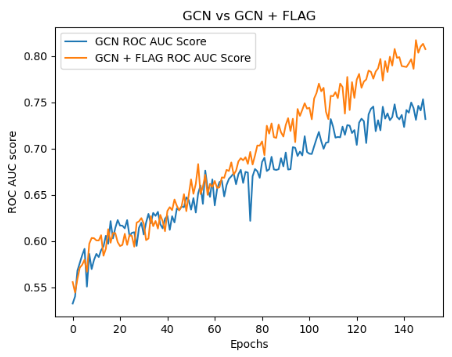}
    \caption{A plot of ROC score for GCN vs the proposed GCN+FLAG sub-architecture.}
    \label{fig:FLAG_auc}
\end{figure} 
As shown in Fig.\ref{fig:FLAG_acc} and Fig.\ref{fig:FLAG_auc}, we observe significant improvement in performance on the use of FLAG. This could be because it preserves the structural integrity of the graphs since only node features are modified. Tox21 is a molecular dataset where random changes in structure may not be realistic. Also, certain constraints are imposed on the perturbations, further improving reliability. FLAG has been found to be effective for discrete features which are commonly encountered in molecular data. In addition to this, FLAG improves generalization, robustness and data diversity, and is computationally efficient with validation accuracy of\textbf{ 70.68\%} and validation AUC-ROC score of\textbf{ 0.806,} both of them greater than the baseline GCN model.

\subsection{Replacing GCNs with GINs}
\subsubsection{Architecture}
\begin{figure}[t]
    \centering
    \includegraphics[width=3in]{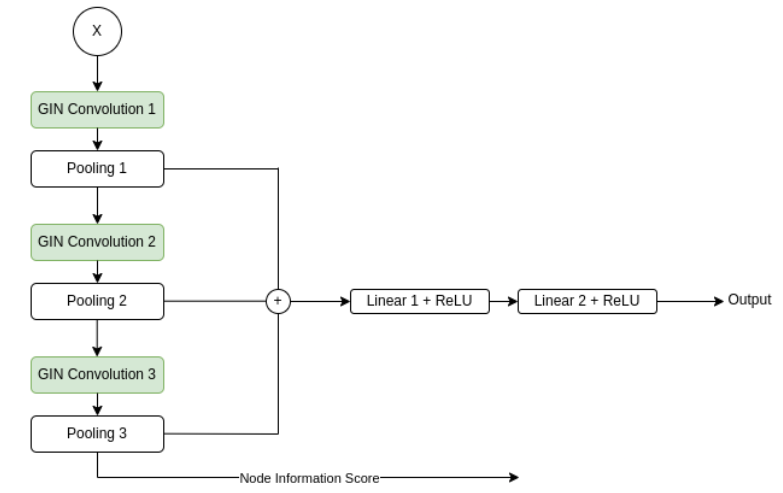}
    \caption{Proposed sub-architecture of GIN based classification model.}
    \label{fig:GIN_arch}
\end{figure} 

The proposed novel architecture (Refer Fig.\ref{fig:GIN_arch}) introduces a modification to the AS-MAML algorithm\cite{asmaml} by replacing the three Graph Convolutional Network (GCN)\cite{gcn} components with three Graph Isomorphism Network (GIN) convolution operators. The GIN operator is anticipated to offer enhanced expressibility, resulting in improved hidden layer embeddings compared to the original GCN-based approach. In the forward() operation of the GIN model, a multi-layer perceptron (MLP) is employed, comprising four hidden layers, each containing ten perceptrons. The output layer, consistent with the GCN model, consists of 128 perceptrons.

\subsubsection{Experimental Results}
\begin{figure}[t]
    \centering
    \includegraphics[width=3in]{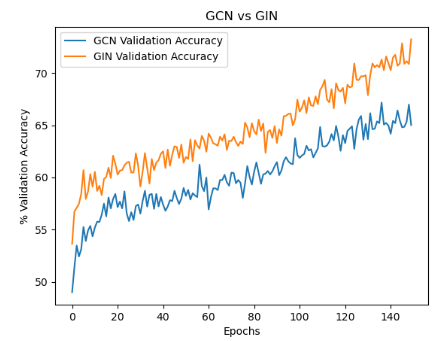}
    \caption{A plot of validation accuracies of the baseline and the proposed GIN sub-architecture.}
    \label{fig:acc}
\end{figure} 
\begin{figure}[t]
    \centering
    \includegraphics[width=3in]{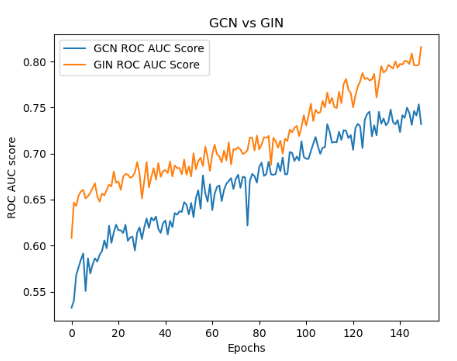}
    \caption{A plot of ROC score for GCN vs the proposed GIN sub-architecture.}
    \label{fig:auc}
\end{figure} 
In the GIN model, we observed significant improvements in validation accuracy and ROC AUC score compared to the baseline GCN model, even at an early stage. The final validation accuracy of \textbf{73.23\%} as shown in Fig.\ref{fig:acc} and ROC score of \textbf{0.816} as shown in Fig.\ref{fig:auc}  can be attributed to the enhanced expressiveness of the Graph Isomorphism Network Operator utilized in GIN. Notably, this improved accuracy is consistently maintained over the course of 150 epochs, suggesting that while GIN may not necessarily provide an advantage in achieving higher-quality training results, it excels at capturing relevant task information with fewer epochs.

\subsection{Enhanced Aggregation and Extraction using Weighted Multi Headed Attention (MHA).}
\subsubsection{Architecture}
\begin{figure}[t]
    \centering
    \includegraphics[width=3.2in]{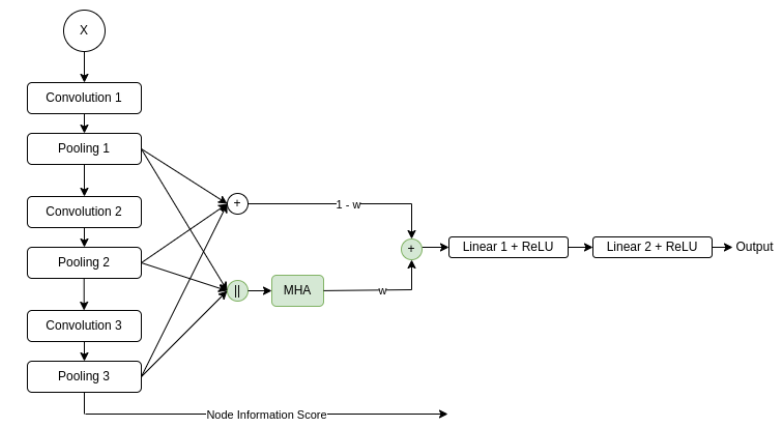}
    \caption{Proposed sub-architecture of GCN+MHA based classification model.}
    \label{fig:MHA_arch}
\end{figure} 
The final setting as shown in Fig.\ref{fig:MHA_arch}, involves adding a Multi Attention Head (MAH) mechanism as an operator in the last part of the AS-MAML model. The baseline GCN carries out a normal aggregation of the outputs of the Relu layers and passes it to a binary classification network as displayed in the baseline figure above which might be unable to extract all necessary information or give weightage to the important ones. By inculcating a MAH layer, we attempt to change this fact and try to make the best out of the convolutions. MAH takes these three values as input to the Key, Value and Query fields to identify patterns of significance. The weight factor \textit{“w”} adds an extra bias to the inclusion of the attention layer while performing regularization. 
\subsubsection{Experimental Reults}
\begin{figure}[t]
    \centering
    \includegraphics[width=3in]{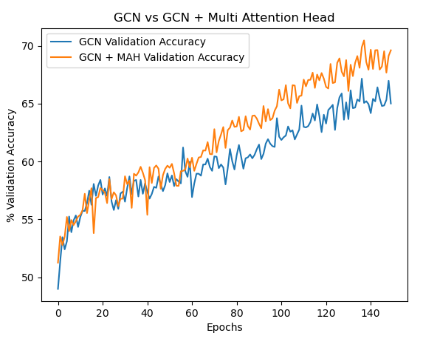}
    \caption{A plot of validation accuracies of the baseline and the proposed GCN+MAH sub-architecture.}
    \label{fig:MHA_acc}
\end{figure} 
\begin{figure}[t]
    \centering
    \includegraphics[width=3in]{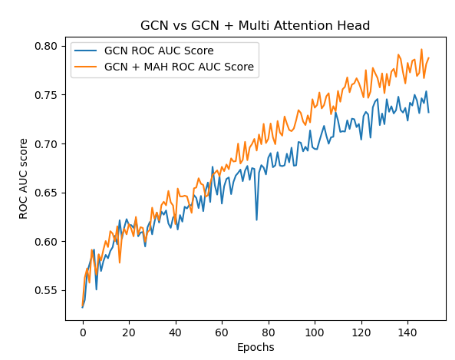}
    \caption{A plot of ROC Score of the baseline and the proposed GCN+MAH sub-architecture.}
    \label{fig:MHA_auc}
\end{figure} 
Upon examination of the graphs presented in Fig.\ref{fig:MHA_acc} and Fig.\ref{fig:MHA_auc}, it becomes evident that the adapted model consistently outperforms the baseline counterpart during the validation phase. The GCN+MAH model achieves a notable validation accuracy of approximately \textbf{69.62\%}, showcasing a significant improvement over the baseline's attainment of 65\%. Additionally, a discernible discrepancy of 0.055 units is observed in the AUC-ROC values, further substantiating the effectiveness of the modified architecture. This enhanced performance of the GCN+MAH model can be attributed to the model's abilty to simultaneously consider multiple attention patterns, enabling it to capture intricate data relationships and patterns more effectively. The experiments have been carried out with the weight factor “w” having a value of 0.4 which is another hyper
parameter that we introduce.

\section{Discussion and Evaluations}\label{sec6}
Within this section, we elaborate on the training regimen and specifications pertinent to Few-Shot Learning (FSL), followed by a comprehensive evaluation of their collective outcomes along with an exhaustive assessment of the best method employed.

\subsection{Programming Environment}
The experiment was conducted on a computer system with the following hardware specifications: an Intel Core i5 8th generation processor, which provides a solid processing capability for running complex algorithms and computations. It is accompanied by a GeForce GTX 1050Ti GPU, which offers parallel processing power and accelerates the training and inference processes for deep learning models. The combination of a capable CPU and GPU allows for efficient execution of the research project.

The operating system used on the computer is Ubuntu, a popular Linux-based operating system known for its stability, security, and compatibility with machine learning frameworks. Specifically, the experiment was conducted with the 6.0.12-76060012-generic kernel.

\begin{table*}[h]
\centering
\caption{Tunable Parameters and Values for FSL setting}
\label{tab:tunable_parameters}
\begin{tabularx}{\textwidth}{|l|X|c|}
\hline
\textbf{Tunable Parameter} & \textbf{Significance} & \textbf{Value in Experiments} \\
\hline
Train Shot & The number of labeled examples from the training set used for adapting the model during the few-shot learning process. & 10 \\
\hline
Validation Shot & The number of labeled examples from the validation set used for fine-tuning or evaluating the model during the few-shot learning process. & 10 \\
\hline
Train Query Set & The set of unlabeled examples from the training set that are used for prediction or evaluation after model adaptation. & 15 \\
\hline
Validation Query Set & The set of unlabeled examples from the validation set that are used for prediction or evaluation after fine-tuning or model evaluation. & 15 \\
\hline
Epochs & Count of how many times the full dataset was run through the model during training. & 150 \\
\hline
Learning Rate (Outer Loop) & A hyperparameter that determines the step size or rate at which the model's parameters are updated during the training process. & 0.001 \\
\hline
Learning Rate (Inner Loop) & A hyperparameter that determines the step size or rate at which the model's parameters are updated during the inner loop training process. & 0.01 \\
\hline
\end{tabularx}
\end{table*}

\subsection{Few Shot Learning Specifications}
In Table \ref{tab:tunable_parameters}, we list some of the common tunable parameters in a few shot learning scenarios and state the settings used for our testing.

\subsection{Comprehensive Analysis}

\begin{table*}[t]
  \caption{The table presents a detailed comparison between the accuracy and ROC values of the proposed methods.}
  \label{tab:model_performance}
  \begin{tabularx}{\textwidth}{|X|X|X|X|X|}
    \hline
    \textbf{Model and Algorithm} & 
    \textbf{Validation Accuracy} &
    \textbf{$\Delta$ Accuracy Score} & 
   \textbf{ ROC-AUC Score} &
    \textbf{$\Delta$ ROC-AUC Score} \\
    \hline
    GCN (Baseline) & 65.02 \% & - & 0.732 & - \\
    \hline
    GCN + FLAG & 70.68  \% & \textbf{+5.66  \%} & 0.806 &\textbf{ +0.074}
    \\
    \hline
    GIN  & 73.23  \% & \textbf{+8.21  \%} & 0.816 & \textbf{ +0.084} \\
    \hline
    GCN + MHA & 69.62  \% & \textbf{+4.6  \%} & 0.787 &\textbf{ +0.055 }
    \\
    \hline
    
  \end{tabularx}
\end{table*}

\begin{figure}[t]
    \centering
    \includegraphics[width=3in]{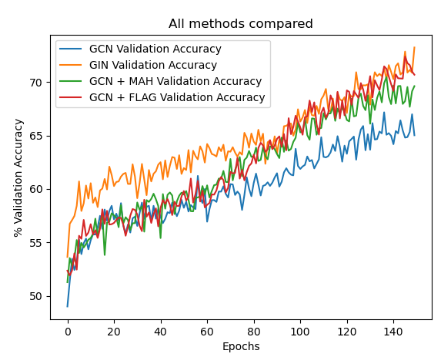}
    \caption{Plot of validation accuracy for baseline and all three proposed sub-architectures.}
    \label{fig:ALL_acc}
\end{figure} 
\begin{figure}[t]
    \centering
    \includegraphics[width=3in]{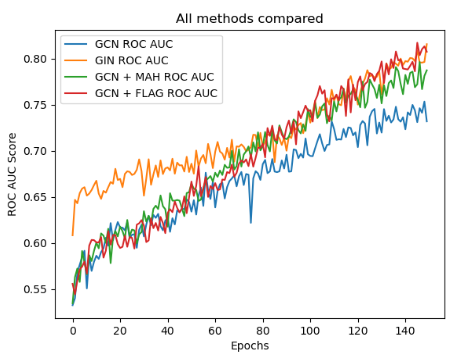}
    \caption{Plot of ROC score for baseline and all three proposed sub-architectures.}
    \label{fig:ALL_auc}
\end{figure} 

The baseline model, GCN, achieves a validation accuracy of 65.02\% and a ROC-AUC score of 0.732. Comparatively, the GCN+FLAG model showcases an improvement with a validation accuracy of 70.68\% (+5.66\%) and a ROC-AUC score of 0.806 (+0.074). The GIN model demonstrates the highest validation accuracy of 73.23\% (+8.21\%) and a ROC-AUC score of 0.816 (+0.084). Additionally, the GCN+MHA model records a validation accuracy of 69.62\% (+4.6\%) and a ROC-AUC score of 0.787 (+0.055). These figures as illustrated in Table \ref{tab:model_performance} and in Fig.\ref{fig:ALL_acc} and Fig.\ref{fig:ALL_auc}, highlight the enhanced expressibility and feature representation ability of GIN operators, attention modules and data augmentation algorithms.  

By adopting these modified architectures, which have the potential to advance the state-of-the-art results, we anticipate achieving improved performance and predictive capabilities in the context of the AS-MAML algorithm for few-shot learning tasks, particularly in toxicological classification of molecular compounds. 

GIN's superiority is attributed to its order-agnostic aggregation operation, which ensures robustness and insensitivity to changes in node positions. In contrast, GCN's performance is influenced by the order of nodes in the neighborhood, making it more sensitive to node ordering. Another aspect contributing to GIN's efficacy is its higher expressiveness compared to GCN. While GCN focuses on local information within fixed neighborhoods, it faces limitations in capturing higher-order graph structures. Conversely, GIN's iterative message passing mechanism enables it to encompass more intricate and global structural patterns, making it more adept at handling complex molecular graphs.

Moreover, GIN's passing of the Weisfeiler-Lehman (WL) test \cite{wltest}, a theoretical measure of a GNN's expressive power, further validates its strength as a graph neural network. The WL test checks whether a GNN can differentiate non-isomorphic graphs with the same initial node labels. GIN's successful performance on this test showcases its stronger representational capacity compared to GCN and other methods.

Nevertheless, it's essential to acknowledge that the efficacy of GNN architectures may vary based on the specific dataset and task, with hyperparameter tuning and data preprocessing also impacting their overall performance.

\section{Future Scope}\label{sec7}
There are several avenues for future work in this research project. Firstly, exploring different variations of the GIN model and incorporating more complex attention mechanisms or utilizing different graph pooling strategies, could potentially enhance its performance. Exploring different meta-learning algorithms or adapting the AS-MAML framework to other graph-based tasks beyond molecular property prediction would expand the applicability and generalizability of the research findings. Finally, conducting experiments on larger and more diverse datasets would provide further insights into the scalability and robustness of the proposed approach.

\section{Conclusion}\label{sec8}
In conclusion, the GIN method presented in this research paper establishes a new benchmark in drug discovery and toxicity prediction using the Tox21 data, exhibiting remarkable improvements of 8.21\% in accuracy and 11.4\% in ROC performance compared to existing GCN methods. Moreover, its exceptional performance in low labeled data scenarios, surpassing all other given methods, underscores its robustness and practicality. This novel approach holds immense promise for researchers and practitioners in pharmaceutical and chemical industries, providing valuable insights and advancements in these fields while inspiring further exploration and adoption of graph neural network-based methodologies for addressing real-world challenges

{
\scriptsize
\bibliographystyle{IEEEtranN}
\bibliography{bibliography}
}
\balance 

\end{document}